\documentclass{article}
\usepackage{spconf,amsmath,graphicx,hyperref}
\usepackage{amsfonts, booktabs}

\title{A long-form single-speaker real-time MRI speech dataset and benchmark}
\name{Sean Foley$^{1,2}$, Jihwan Lee$^{1}$, Kevin Huang$^{1}$, Xuan Shi$^{1}$, \\\em{Yoonjeong Lee$^{1}$, Louis Goldstein$^{2}$, Shrikanth Narayanan$^{1,2}$}}
\address{$^1$Signal Analysis and Interpretation Lab, University of Southern California \\
$^2$Department of Linguistics, University of Southern California}
\begin{document}
%
\maketitle
\begin{abstract}
We release the USC Long Single-Speaker (LSS) dataset containing real-time MRI video of the vocal tract dynamics and simultaneous audio obtained during speech production. This unique dataset contains roughly one hour of video and audio data from a single native speaker of American English, making it one of the longer publicly available single-speaker datasets of real-time MRI speech data. Along with the articulatory and acoustic raw data, we release derived representations of the data that are suitable for a range of downstream tasks. This includes video cropped to the vocal tract region, sentence-level splits of the data, restored and denoised audio, and regions-of-interest timeseries. We also benchmark this dataset on articulatory synthesis and phoneme recognition tasks, providing baseline performance for these tasks on this dataset which future research can aim to improve upon. 
Dataset website: \url{https://sail.usc.edu/span/single_spk}
\end{abstract}
%
\section{Introduction}
\label{sec:intro}
Speech production requires the formation of constrictions in the vocal tract, employing multiple articulators in a synergistic fashion to achieve such constrictions efficiently and effectively~\cite{turvey1977preliminaries,browman1992articulatory}. The crucial active articulators utilized during speech production include minimally the lips, jaw, tongue, velum, and larynx. While a range of methods exist to capture these various articulators during speech production, real-time (rt)MRI bestows the most in-depth view of the vocal tract, providing a full midsagittal view of the vocal tract including the larynx, pharynx and velum which are usually difficult to image non-invasively. 
Naturally, rtMRI comes with the downside of being expensive to acquire, run and maintain, making publicly released datasets enormously beneficial to the speech science and technology community.
\par
A number of previous rtMRI speech datasets have been publicly released. The USC-TIMIT dataset~\cite{narayanan2014real} contains rtMRI video and audio from 10 speakers, 5 male and 5 female, producing 460 phonetically rich sentences, captured at 23 FPS. For each speaker, there is roughly 37 minutes of speech. The USC 75-Speaker dataset~\cite{lim2021multispeaker} contains a mix of read and spontaneous speech from 75 speakers, including speakers of American English, Indian English, and native speakers of other languages, e.g., Chinese. There is approximately 17 minutes of speech for every speaker and the MRI video was reconstructed at 86 FPS. The dataset released here extends upon these earlier datasets by focusing on capturing more data for an individual speaker.


\par
Speech data collected using rtMRI has been applied to a range of speech processing tasks and phonetic analyses, such as articulatory synthesis and speech inversion. While electromagnetic articulography and ultrasound articulatory data have also been used in these tasks, they typically require additional acoustic features to capture nasality and voicing~\cite{cho2024articulatory}. rtMRI-based approaches can achieve similar performance directly without any additions~\cite{wu2023deep}. Self-supervised representation learning from articulatory data \cite{lian2023articulatory} has been used in brain-computer interfaces \cite{metzger2023high} and automated pronunciation assessment \cite{lian2024ssdm}, with rtMRI datasets being deployed in these developments \cite{lian2023articulatory}. Lastly, phoneme recognition from rtMRI and its corresponding audio have allowed for multimodal modeling and representation learning~\cite{shi2024direct, foley2025towards}. 


We release the USC Long Single-Speaker (LSS) dataset, which contains roughly one hour of rtMRI speech data from a single native speaker of American English. This offers considerably more single-speaker speech data, particularly of spontaneous speech, than previously released datasets. As such, we anticipate that this dataset can aid in furthering model development in tasks such as articulatory synthesis and speech inversion that are challenging to perform across multiple speakers. 


\section{USC Long Single-Speaker (LSS) Dataset}
\label{sec:corpus}

\subsection{Overview}

The USC LSS dataset contains nearly one hour of speech from a single speaker collected using rtMRI. To our knowledge, this makes this dataset the longest publicly available single-speaker rtMRI dataset. This dataset contains rtMRI video of the vocal tract during speech production and synchronized audio. Furthermore, while previous datasets typically only release the original audio and video, we include also several representations derived from the original data that may be more suitable for various forms of research, including video cropped to the vocal tract region, two forms of processed audio - denoised and restored, region-of-interest timeseries, and sentence-level splits of the data. 


\begin{table}[t]
\centering
\begin{tabular}{lccc}
\toprule
\textbf{Dataset} & \textbf{Read (min)} & \textbf{Spont. (min)} & \textbf{FPS} \\
\midrule
USC TIMIT \cite{narayanan2014real}       & $\sim37$ & 0       & 23 \\
USC 75 Speaker \cite{lim2021multispeaker} & $\sim12$ & $\sim5$ & 86 \\
\midrule
USC LSS                                   & 37       & 17      & 99 \\
\bottomrule
\end{tabular}
\caption{Comparison of longest single-speaker durations for read and spontaneous speech and FPS for previous datasets versus the current dataset.}
\label{tab:comp}
\end{table}

\subsection{Data Acquisition}

\begin{figure}[t]
    \centering
    \begin{minipage}[b]{0.48\linewidth}
        \centering
        \includegraphics[width=\linewidth]{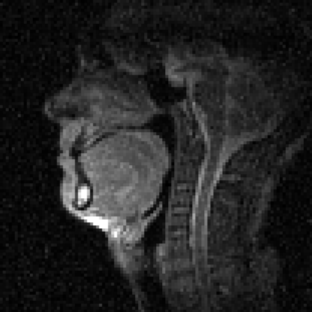}
    \end{minipage}%
    \hfill
    \begin{minipage}[b]{0.48\linewidth}
        \centering
        \includegraphics[width=\linewidth]{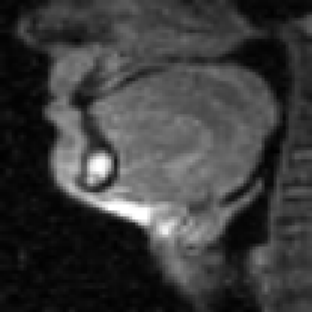}
    \end{minipage}
    \caption{Example frame from the original video (left) and the
video cropped to the vocal tract region (right)}
    \label{fig:frame}
\end{figure}

The corpus contains speech data from one male native speaker of American English aged 32 producing a combination of read and spontaneous speech. The read speech includes the 460 sentences used in the USC TIMIT corpus and two repetitions of the grandfather, rainbow, and northwind passages used in the USC 75-Speaker dataset. The spontaneous speech includes two repetitions of five picture descriptions and prompts on topics including food, travel, music, and movies. The vocal tract of the speaker was imaged in midsagittal orientation using a 0.55T MRI scanner with a custom upper airway receiver coil \cite{munoz2023evaluation}. Acquired data was reconstructed at a frame rate of 99 frames/sec (FPS), see \cite{kumar24b_interspeech} for details. Audio was collected at 16 kHz. In total, 54 minutes of data was collected over 71 scans, with 37 minutes of read speech and 17 minutes of spontaneous speech (See Table \ref{tab:comp} for comparison with previous datasets). First-pass phoneme alignments were extracted using the Montreal Forced Aligner (MFA)~\cite{mcauliffe2017montreal} and manually corrected by a phonetician in Praat. In addition to the original video, we also release video cropped to the vocal tract region, with the upper hard palate, larynx, and pharyngeal wall serving as boundaries (See Figure \ref{fig:frame}), allowing for only essential speech-relevant information in the video frame. 

\subsection{Restored \& Denoised Audio}
\label{sec:rest}
Along with the original audio, we provide a restored version of audio recordings by a recent speech restoration model, Miipher~\cite{koizumi2023miipher}\footnote{\url{https://github.com/Wataru-Nakata/miipher}} with reduced background MRI device noise.
Unlike typical speech denoising models where only speech acoustics is used as input, Miipher takes text representation as input along with the speech acoustics, maximizing usage of text information for inaudible parts, outputting a cleaner version of the input speech. We recommend exercising caution when this version is used as some of the restored parts of originally inaudible segments may not match to the corresponding articulatory kinematics. This restored version may be useful only when such precise correspondence between speech acoustics and articulatory kinematics is not strictly required. We also include audio denoised using the Denoiser model \cite{defossez2020real}. 

\subsection{Sentence-level Split}
\label{sec:sent}
In addition to the full original files, we include sentence-level video and audio files, including the original, denoised, and restored audio. Audio transcripts were created semi-automatically using the original stimuli as a baseline for the scripted speech and Whisper-large \cite{radford2023robust} ASR as a baseline for the spontaneous speech. These baselines were adjusted manually based on the audio. From these transcripts, a custom script was used to split the audio and video files using punctuation, resulting in a total of 684 sentences. We further split these sentences into train, validation, and test (0.85/0.05/0.1) sets to be employed in any downstream task. 

\subsection{Articulatory Region-of-interest (ROI) Timeseries}

\begin{figure}[t]
    \centering
    \includegraphics[width=0.55\linewidth]{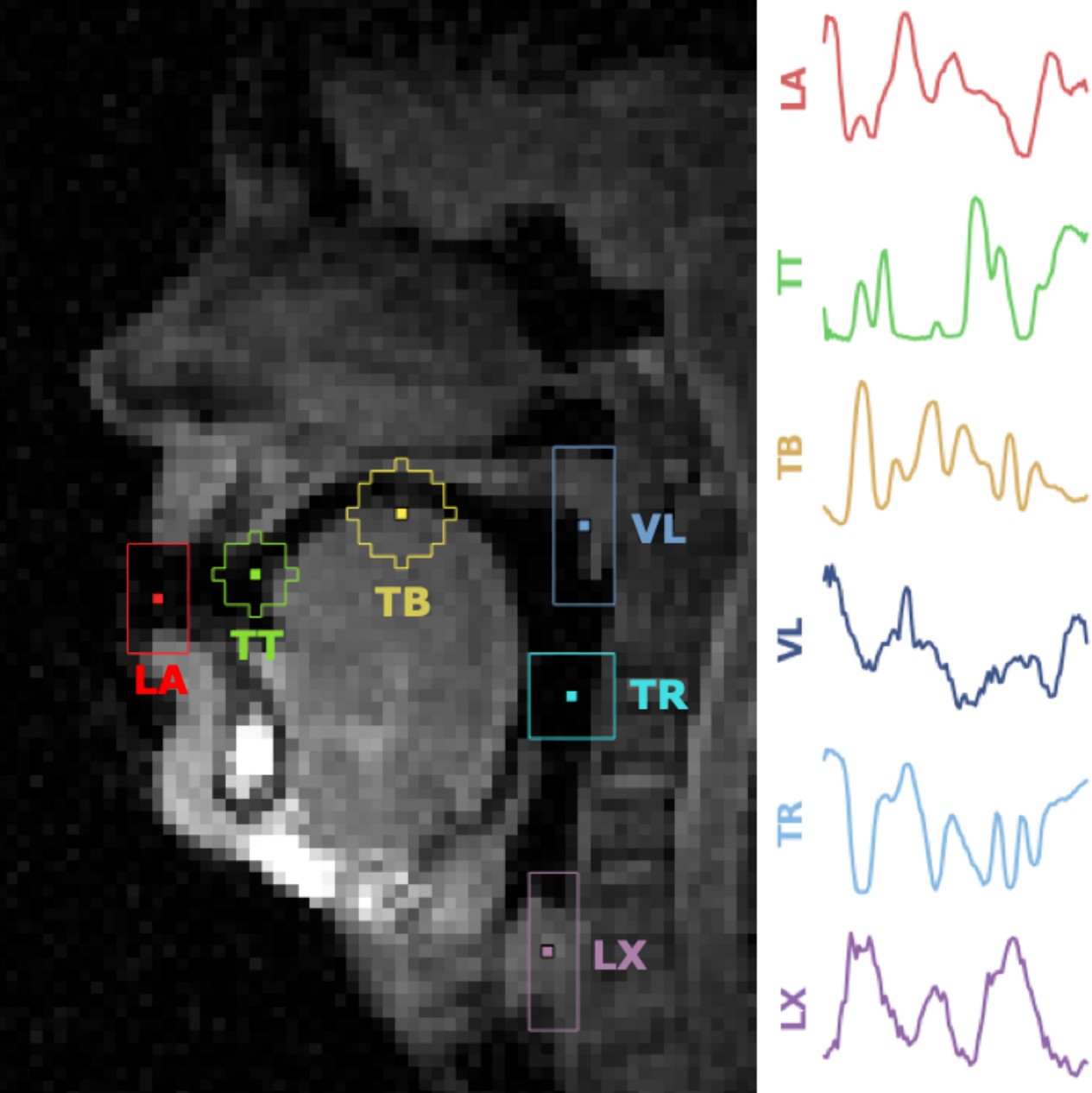}
    \caption{Example frame with regions of interest (ROI) marked (left) and their corresponding timeseries (right).}
    \label{fig:roi}
\end{figure}

As shown in Figure~\ref{fig:roi}, six interpretable, representative regions of interest (ROIs) from the vocal tract that are minimally required to represent speech in English were chosen as follows~\cite{browman1992articulatory}: Lip Aperture (LA), Tongue Tip (TT), Tongue Body (TB), Velum (VL), Tongue Root (TR), and Larynx (LX). Each region is manually annotated by speech researchers to best capture the movement within each region, using the VocalTract ROI Toolbox\footnote{\url{https://github.com/reedblaylock/VocalTract-ROI-Toolbox}}. For each ROI, pixel intensity is calculated as a proxy measure for constriction degree. 
We include the ROI timeseries data for researchers to explore speech processing tasks within this low-dimensional space.

\section{Benchmark Baselines}
\label{sec:bench}
Our aim here is not to optimize task performance but to establish reproducible baselines that future work can build on. These results highlight what can be achieved `out of the box' with common architectures when trained on the USC LSS dataset.

\subsection{Articulatory Synthesis}

\par\noindent\textbf{Model}. Similar to previous work \cite{wu2023deep, wu2024deep}, we employ a neural vocoder to synthesize waveforms from articulatory features. Specifically, we use the HiFi-GAN model~\cite{kong2020hifi}, which is composed of a Generator and two Discriminator networks, a Multi-Period Discriminator (MPD) and Multi-Scale Discriminator (MSD). For our experiments, we use a pretrained Generator, MPD, and MSD trained on the VCTK~\cite{vctk}, Librispeech~\cite{panayotov2015librispeech}, and LJSpeech~\cite{ljspeech17} datasets. We freeze the last of four Generator blocks and both the MPD and MSD to prevent the discriminators from overpowering the Generator.   
\par\noindent\textbf{Preprocessing}. We experiment with using the denoised audio compared to the restored audio described in Section \ref{sec:rest}. All audio had a sampling rate of 16 kHz. The video frames were $z$-scored and resized to $128 \times 128$ which resulted in better performance compared to the original frame size. Videos were reshaped to [$H\times W, t$], where $t$ is the number of frames, and loaded in grayscale to maintain the 1D convolutions used in the Generator.
\par\noindent\textbf{Implementation}. The input size to the Generator was changed to  $128 \times 128$, while all other dimensions were unchanged. We used a hop size of 162, upsample rates of [6, 3, 3, 3], and upsample kernels of [12, 6, 6, 6] across the four blocks. A batch size of 2 and a learning rate of 1e-3 were used during training. Other hyperparameters were the same as the V1 pretrained HiFi-GAN. We used the sentence-level splits for train, validation and test sets and report the results from the held-out test set. 
\par\noindent\textbf{Results}. The articulatory synthesis results are shown in Table \ref{tab:synth}. Interestingly, the model trained on the restored speech outperforms the model trained on denoised speech on Character Error Rate (CER) and Word Error Rate (WER), but the latter performers better on Mel-Cepstral Distance (MCD). This may be attributable to some aspect of the restored speech itself, rather than actual model performance. Overall, the subjective performances for the restored audio model outperform previous studies that have attempted direct rtMRI articulatory features to speech \cite{wu2023deep, wu2024deep}, while the denoised audio model is on par with this previous work. 
\par
\begin{figure}
    \centering
    \includegraphics[width=\linewidth]{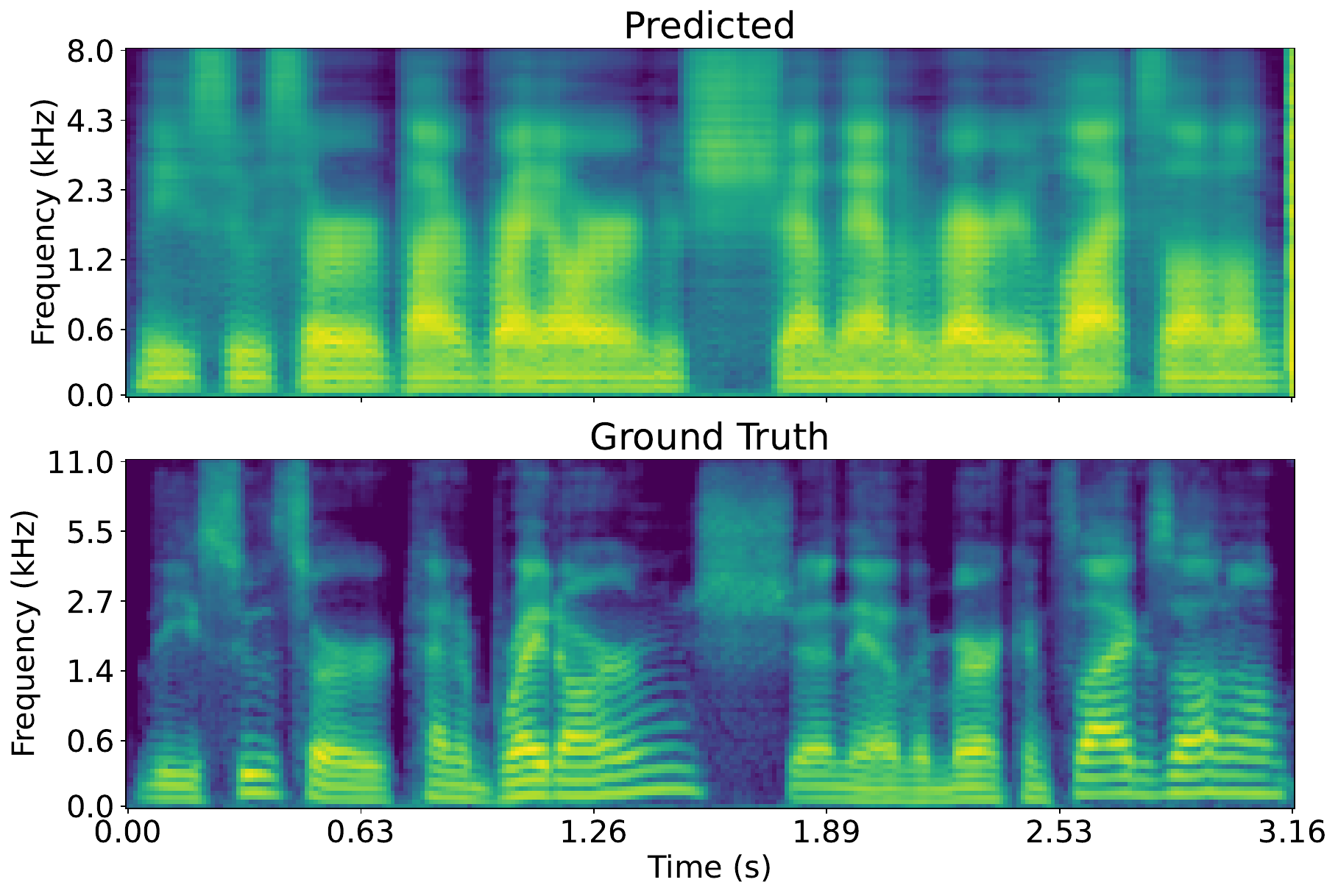}
    \caption{Example predicted (top) and ground truth (bottom) spectrograms from the restored speech model.}
    \label{fig:spec_comp}
\end{figure}

An example of predicted speech from the model trained on restored speech can be seen in Figure \ref{fig:spec_comp}. In comparison to the ground truth spectrogram (bottom), the predicted spectrogram (top) clearly lacks the fine resonance structure typical of natural speech. While general syllabic and resonance structure can be seen in the predicted speech, it exhibits rather thick bands that approximate the formant structure. This still allows for the speech to sound intelligible, but renders it quite unnatural. 

\begin{table}[t]
\centering
\begin{tabular}{lccc}
\toprule
\textbf{Audio} & \textbf{CER (\%)} $\downarrow$ & \textbf{WER (\%)} $\downarrow$ & \textbf{MCD} $\downarrow$ \\
\midrule
Denoised & 31.3 $\pm$ 6.0 & 52.6 $\pm$ 10.1 & 4.3 $\pm$ 0.1 \\
Restored & 24.3 $\pm$ 4.6 & 45.0 $\pm$ 8.2 & 4.8 $\pm$ 0.2 \\
\bottomrule
\end{tabular}
\caption{Speech synthesis quality results for all models trained in the current study with utterance-wise averages and 95\% confidence intervals.}
\label{tab:synth}
\end{table}

\subsection{Phoneme Recognition}

We follow the general set up reported in \cite{foley2025towards}, with the goal of comparing differences in phoneme recognition performance as a function of input modality, i.e. comparing unimodal audio and video models to a multimodal model, see \cite{foley2025towards} for further details. 
\par\noindent\textbf{Model}. The primary architecture is a Conformer\cite{gulati2020Conformer}. The output of the Conformer is decoded by a single LSTM layer, with a final linear layer used for prediction. Unimodal input to the Conformer consists of either acoustic features or video features. For the multimodal model, the audio and video features are concatenated along the temporal dimension.
All models are trained using the connectionist temporal classification (CTC) loss \cite{ctc-loss}. As a baseline, we performed zero-shot phoneme recognition on the audio using Wav2Vec2Phoneme, accessed via Hugging Face.
\par\noindent \textbf{Preprocessing}. We use the sentence-level splits described in Section \ref{sec:sent} and the denoised audio for training. Audio features were extracted from the $9^{th}$ layer of the pre-trained base WavLM model~\cite{chen2022wavlm} and video features comprised of the classification (CLS) token for each frame from the last hidden layer of a fine-tuned base ViT model~\cite{alexey2020image}.
\par\noindent\textbf{Implementation}.
The input dimension to the Conformer was set to 768 for the unimodal models and 2*768 for the multimodal model, with a feed-forward dimension of 256. The number of attention heads was set at 4, a kernel size of 31, and dropout set at 0.3. The number of Conformer layers was set to 3. The LSTM latent size was 128, which served as the input size for the final linear layer. For all models, the Adam optimizer was used, with a batch size of 16 and a learning rate of 1e-4, which was decayed by 0.9 every 20 epochs.

\par\noindent \textbf{Results}. The Phoneme Error Rate (PER) results are shown in Table \ref{tab:per}. The unimodal audio model performs best, outperforming the other two models and the baseline. The video only model performs worst, with a PER of only 0.50. Interestingly, the multimodal model performs worse than the audio only model, suggesting simple concatenation is not an ideal method for combining features across modalities in this context. We encourage future work to explore alternative approaches to multimodal learning, such as cross-modal attention \cite{lu2016hierarchical} and multimodal mixture-of-experts \cite{mustafa2022multimodal}. 
\par
\begin{figure}[t]
    \centering
    \includegraphics[width=\linewidth]{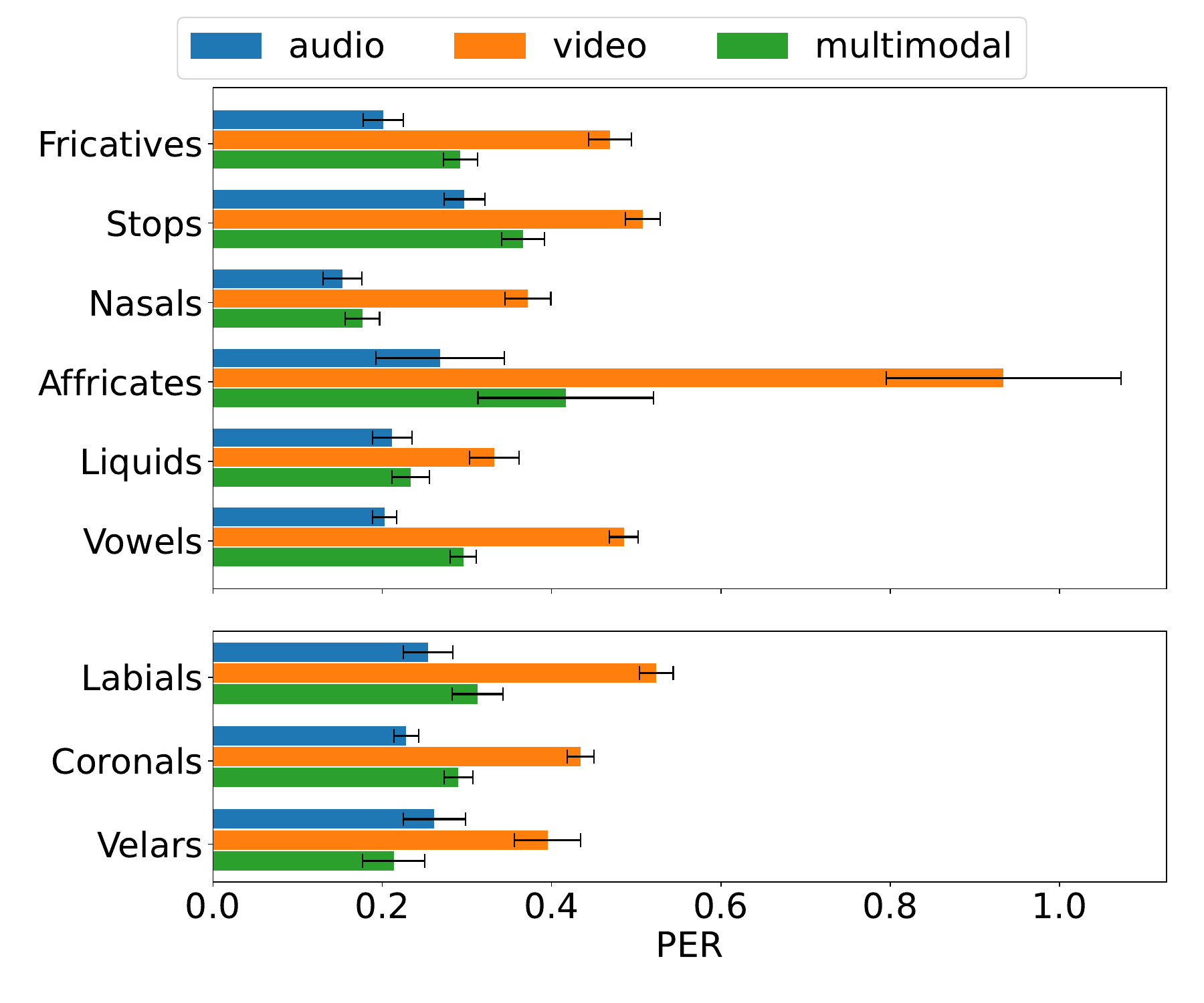}
    \caption{PER results by phonetic class for unimodal and multimodal models, grouped by manner (above) and place of articulation (below).}
    \label{fig:per}
\end{figure}

Figure  \ref{fig:per} shows PER per phonetic manner and place classes across the three models. Noticeably, multimodal performs much worse than audio only in cases when articulatory distinctions are subtle in rtMRI video, as in stops and fricatives, or highly variable, as in vowels \cite{whalen2018variability}. Multimodal performance is best when both the acoustic and articulatory signatures are distinct, as in liquids and nasals. The multimodal model actually performs best of the three on velars, which are typically easily distinguishable in articulation from labials and coronals and have distinct formant transitions. 

\begin{table}[t]
\centering
\begin{tabular}{lc}
\toprule
\textbf{Model} & \textbf{PER} $\downarrow$\\
\midrule
Wav2Vec2Phoneme & 0.39 $\pm$ 0.03 \\ \midrule
Audio           & 0.22 $\pm$ 0.03\\
Video           & 0.50 $\pm$ 0.02\\
Multimodal      & 0.30 $\pm$ 0.03\\
\bottomrule
\end{tabular}
\caption{Utterance-wise average PER with 95\% confidence intervals for models in the current study (bottom) compared to the baseline model (top).}
\label{tab:per}
\end{table}

\section{Research Applications}
\label{sec:research}
The USC LSS dataset can be used for a range of speech processing tasks and phonetic analyses.  rtMRI-based approaches to tasks like speech inversion and articulatory synthesis have been limited by the availability of long single-speaker data \cite{wu2023deep}. This dataset can aid in developing models that learn mappings between acoustics and articulation. Additionally, vocal tract contours can serve as robust articulatory representations for many tasks \cite{bresch2008region,wu2023deep,shi202575}, but automated methods for contour extraction from 0.55T magnets generating 99 FPS video are underdeveloped. This dataset can help further these efforts. 
\par
For phoneticians, the USC LSS dataset provides ample opportunity for studying within-speaker production, particularly concerning variability and uniformity \cite{chodroff2022uniformity}. Speakers are known to exhibit considerable variability in the production of a given segment \cite{harperindividual}, while also demonstrating a degree of uniformity across segments in terms of ``reusing" certain articulatory gestures \cite{faytak2022place,chodroff2022uniformity}. While the development of corpus phonetics has allowed for data-driven research in these areas, the extensive paired audio and articulatory data presented here can surely aid in understanding within-speaker production.

\section{Acknowledgments}
This work was supported by NIH grant T32 DC009975 and NSF 2311676.
\bibliographystyle{IEEEbib}
\footnotesize
\bibliography{My_Library}

\end{document}